\documentclass{article}

\PassOptionsToPackage{numbers, compress}{natbib}

\usepackage[final]{neurips_2021}

\usepackage[utf8]{inputenc} 
\usepackage[T1]{fontenc}    
\usepackage{hyperref}       
\usepackage{url}            
\usepackage{booktabs}       
\usepackage{amsfonts}       
\usepackage{nicefrac}       
\usepackage{microtype}      
\usepackage{xcolor}         
\usepackage{mathtools}
\usepackage{amssymb}
\usepackage{tikz}
\usetikzlibrary{bayesnet}
\usepackage{subcaption}
\newcommand{\KL}[2]{\mbox{KL}\left[ #1 \,\|\, #2 \right]}

\title{Targeted Neural Dynamical Modeling}

\author{%
  Cole Hurwitz \\
  School of Informatics\\
  University of Edinburgh\\
  Edinburgh, Scotland, EH8 9AB \\
  \texttt{colehurwitz@gmail.com} \\
   \And
   Akash Srivastava \\
   MIT-IBM Watson AI Lab \\
   Cambridge, MA 02142,  \\
   \texttt{Akash.Srivastava@ibm.com} \\
   \AND
   Kai Xu \\
   School of Informatics\\
   University of Edinburgh\\
   Edinburgh, Scotland, EH8 9AB \\
   \texttt{xukai921110@gmail.com} \\
   \And
   Justin Jude\\
   School of Informatics\\
   University of Edinburgh\\
   Edinburgh, Scotland, EH8 9AB \\
   \texttt{justin.jude@ed.ac.uk} \\
   \And
   Matthew G. Perich \\
   Icahn School of Medicine at Mount Sinai \\
   New York, NY 10029 \\
   \texttt{mperich@gmail.com} \\
   \And
   Lee E. Miller \\
   Feinberg School of Medicine \\
   Northwestern \\
   Chicago, IL 60611\\
   \texttt{lm@northwestern.edu} \\
   \And
   Matthias H. Hennig \\
   School of Informatics\\
   University of Edinburgh\\
   Edinburgh, Scotland, EH8 9AB \\
   \texttt{m.hennig@ed.ac.uk} \\
}

\begin{document}

\maketitle

\begin{abstract}
Latent dynamics models have emerged as powerful tools for modeling and interpreting neural population activity. Recently, there has been a focus on incorporating simultaneously measured behaviour into these models to further disentangle sources of neural variability in their latent space. These approaches, however, are limited in their ability to capture the underlying neural dynamics (e.g. linear) and in their ability to relate the learned dynamics back to the observed behaviour (e.g. no time lag). To this end, we introduce Targeted Neural Dynamical Modeling (TNDM), a nonlinear state-space model that jointly models the neural activity and external behavioural variables. TNDM decomposes neural dynamics into behaviourally relevant and behaviourally irrelevant dynamics; the relevant dynamics are used to reconstruct the behaviour through a flexible linear decoder and both sets of dynamics are used to reconstruct the neural activity through a linear decoder with no time lag. We implement TNDM as a sequential variational autoencoder and validate it on simulated recordings and recordings taken from the premotor and motor cortex of a monkey performing a center-out reaching task. We show that TNDM is able to learn low-dimensional latent dynamics that are highly predictive of behaviour without sacrificing its fit to the neural data.
\end{abstract}

\section{Introduction}
\label{sec:intro}
Recent progress in high-density, microelectrode array technology now allows for recording from hundreds to thousands of neurons with the precision of single spikes \cite{jun2017fully}. Despite the apparent high dimensionality of these datasets, neural activity is often surprisingly well-explained by low-dimensional latent dynamics \cite{churchlandNeuralPopulationDynamics2012, sadtler2014neural, elsayed2017structure, gallego2017neural}. Extracting these dynamics from single trials is crucial for understanding how neural activity relates to a behavioural task or stimulus \cite{pandarinath2018inferring}.

Latent variable models (LVMs) are a natural choice for capturing low-dimensional structure from neural activity as they can learn to map a few latent variables to arbitrarily complicated response structure in the activity. Already, there exist a number of LVMs that have been successfully applied to neural data ranging from simple non-temporal models such as principal components analysis (PCA) \cite{cunningham2014dimensionality} to complex state-space models such as LFADS \cite{pandarinath2018inferring}. In these models, the goal is to learn a set of latent factors that best explain neural variability. As such, there is no guarantee that the different sources of variability present in the population activity will be disentangled in the latent space (e.g. behaviour, arousal, thirst, etc.) \cite{saniModelingBehaviorallyRelevant2020, hurwitz2021building}. 

To better partition sources of neural variability in the latent space, some LVMs have been developed that incorporate an external behaviour into the generative process \cite{kobak2016demixed, perich2020motor, zhou2020learning}. These methods, however, do not model temporal dependencies between the latent states. Recently, a novel state-space model termed preferential subspace identification (PSID) was developed that jointly models neural activity and behaviour with a shared set of dynamics \cite{saniModelingBehaviorallyRelevant2020}.
When applied to neural activity recorded in the premotor cortex (PMd) and primary motor cortex (M1) of a monkey during a 3D reaching task, PSID was shown to extract latent factors that were more predictive of behaviour than the factors extracted by other approaches. Despite the strength and simplicity of this approach, it suffers from two main drawbacks. First, PSID is a linear state-space model and cannot capture the nonlinear dynamics which are thought to underlie phenomena such as rhythmic motor patterns \cite{russo2020neural, hall2014common} or decision making \cite{rabinovich2008transient}. Second, PSID assumes that behaviourally relevant dynamics explain both the neural activity and behaviour with no time lag. This limits the ability of PSID to capture more complex temporal relationships between the latent dynamics and the behaviour.


In this work, we introduce Targeted Neural Dynamical Modeling (TNDM), a nonlinear state-space model that jointly models neural activity and behaviour. Similarly to PSID, TNDM decomposes neural activity into behaviourally relevant and behaviourally irrelevant dynamics and uses the relevant dynamics to reconstruct the behaviour and both sets of dynamics to reconstruct the neural activity. Unlike PSID, TNDM does not constrain the latent dynamics at each time step to explain behaviour at each time step and instead allows for any linear relationship (constrained to be causal in time) between the relevant dynamics and the behaviour of interest. We further encourage partitioning of the latent dynamics by imposing a disentanglement penalty on the distributions of the initial conditions of the relevant and irrelevant dynamics. To perform efficient inference of the underlying nonlinear dynamics, TNDM is implemented as a sequential variational autoencoder (VAE) \cite{kingma2013auto, sussillo2016lfads}\footnote{The code for running and evaluating TNDM on real data can be found at \href{https://github.com/HennigLab/tndm_paper}{https://github.com/HennigLab/tndm\_paper}. We also provide a Tensorflow2 re-implemention of TNDM at \href{https://github.com/HennigLab/tndm}{https://github.com/HennigLab/tndm}. It is important to note that all reported results for the \textit{real} datasets use the old model and not the re-implementation. For the \textit{synthetic} dataset results, we use the re-implementation.}. We compare TNDM to PSID and to LFADS, a nonlinear state-space model that only models neural activity, to illustrate that TNDM extracts more behaviourally relevant dynamics without sacrificing its fit to the neural data. We validate TNDM on simulated recordings and neural population recordings taken from the premotor and motor cortex of a monkey during a center-out reaching task. In this analysis, we find that the behaviourally relevant dynamics revealed by TNDM are lower dimensional than those of other methods while being more predictive of behaviour.






\section{Background/Related work}
\label{sec:background}
\paragraph{Notatation.} Let $x \in \mathbb{N}^{N\times T}$ be the observed spike counts and let $y \in \mathbb{R}^{B\times T}$ be the observed behaviour during a single-trial.\footnote{In this work, we assume that behaviour is temporal and has the same time length as recorded neural activity (e.g. hand position). TNDM can be extended to discrete/non-temporal behaviours (e.g. reach direction).} We define the unobserved latent factors in a single trial as  $z \in \mathbb{R}^{M\times T}$ where $M < N$. 
For TNDM, as with PSID, it is important to distinguish between behaviourally relevant $z_r$ and behaviourally irrelevant $z_i$ latent factors. The behaviourally relevant latent factors $z_r$ summarize the variability in the neural activity associated with the observed behaviour while the behaviourally irrelevant latent factors $z_i$ explain everything else in the neural data (Figure \ref{fig:tndm_prob}). We assume that each of the unobserved, single-trial factors can be partitioned into these relevant and irrelevant factors $z \coloneqq \{z_r, z_i\}$. 

\paragraph{State-space models for neural data.} There are a number of state-space models that have been developed and applied to neural population activity. The expressivity of these models range from simple linear dynamical models \cite{smith2003estimating, buesing2013spectral, pfau2013robust} to more complex nonlinear models where the latent dynamics are parameterized by recurrent neural networks (RNNs) \cite{pandarinath2018inferring, she2020neural}. For this work, there are two state-space models that are most relevant: LFADS and PSID. 

LFADS, or latent factor analysis via dynamical systems, is a state-of-the-art nonlinear state-space model for neural data. In LFADS, the latent dynamics are generated by sampling high-dimensional initial conditions $g_0$ from some distribution $p_{g_0}$ and then evolving $g_0$ with a deterministic RNN $f_\theta$. A linear mapping $W_z$ is then applied to the high-dimensional dynamics $g_t$ to transform them into the low-dimensional ‘dynamical’ factors $z$. These dynamical factors are transformed into spike counts by mapping each time point to a rate parameter $r$ of a Poisson distribution using a weight matrix $W_r$ followed by an exponential nonlinearity. The generative process is defined as: $g_0 \sim p_{g_0}, g_t = f_\theta(g_{t-1}), z_t = W_z(g_t), r_t = \exp(W_r(z_t)), x_t \sim \text{Poisson}(x_t | r_t)$. The initial conditions $g_0$ are inferred from $x$ with an RNN encoder network $q_\phi$. Utilizing the reparameterization trick, the model is trained using gradient descent and by optimizing the evidence lower-bound (ELBO) of the marginal log-likelihood. While LFADS provides an excellent fit to the neural data, it inevitably mixes different sources of neural variability in the latent dynamics $z$ as there is no constraints imposed to disentangle these dynamics.

PSID is a linear dynamical model that partitions the latent dynamics into behaviourally relevant and behaviourally irrelevant dynamics $z \coloneqq \{z_i, z_r\}$. The dynamical evolution of $z$ is defined by a transition matrix $A$ along with a Gaussian noise term $w_z$. $z$ is transformed into the observed firing rates $x$ by mapping each time point to the mean of a Gaussian distribution with a weight matrix $W_x$ and noise term $w_x$. The behaviourally relevant dynamics $z_r$ at each time point are transformed into the observed behaviour $y$ with a weight matrix $W_y$. The state space model for PSID is then defined as: $z_t \sim \mathcal{N}(z_t | A(z_{t-1}), w_z), x_t \sim \mathcal{N}(x | W_x(z_t), w_x), y_t = W_y(z_r)$. PSID uses a novel two-stage subspace identification approach to learn the parameters of their model. In the first stage, PSID extracts the behaviourally relevant dynamics through an orthogonal projection of future behaviour onto past neural activity. The irrelevant dynamics are then extracted through an additional orthogonal projection of residual neural activity onto past neural activity. In comparison to LFADS, PSID was shown to extract latent states that are better able to predict behaviour when using a Kalman filter. Despite the analytical simplicity of this approach, it suffers from a few drawbacks. First, it can only model linear dynamics which may not provide a good fit to nonlinear activity patterns or behaviours (e.g. multi-directional reaches). Second, the relevant dynamics at each time step ${z_r}_t$ must be mapped one-to-one to the behaviour ${y}_t$ during training (i.e. no time lag). This imposes a strong structural constraint on the relevant dynamics which hampers their ability to explain neural variability.





\section{Model}
\label{sec:model}
\begin{figure}
\begin{subfigure}{.2\textwidth}
  \centering
  \tikz{
 \node[obs] (x) {$x$};%
 \node[obs, right=of x] (y) {$y$};%
 \node[det,above=of x,yshift=-.5cm,xshift=0cm] (z_{i}) {$z_{i}$}; %
 \node[det,above=of y,yshift=-.5cm,xshift=0cm] (z_{r})
 {$z_{r}$}; %
 \node[latent,above=of z_{i},yshift=-.5cm, xshift=0cm] ({g_0}_i) {${g_0}_i$}; %
 \node[latent,above=of z_{r},yshift=-.5cm,xshift=0cm] ({g_0}_r)
 {${g_0}_r$}; %
 \plate [inner sep=.2cm,yshift=.1cm] {plate0}
 {(x)(y)(z_{i})(z_{r})({g_0}_i)({g_0}_r)} {$K$}; %
 \edge {z_{i},z_{r}} {x}  
 \edge {z_{r}} {y}
 \edge {{g_0}_i} {z_{i}} 
 \edge {{g_0}_r} {z_{r}} 
 }
  \centering
  \captionsetup{justification=centering}
  \caption{TNDM \\ graphical model}
  \label{fig:tndm_prob}
\end{subfigure}
\begin{subfigure}{.80\textwidth}
  \includegraphics[scale=.35]{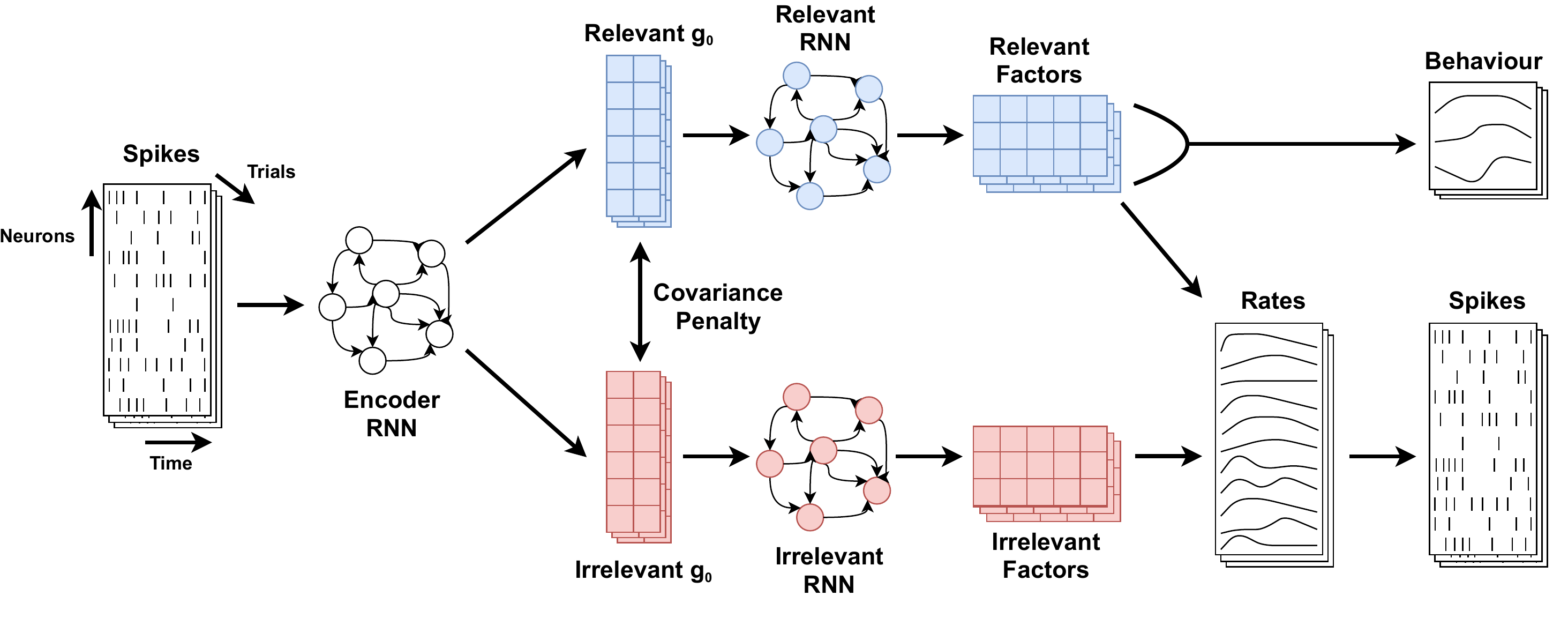}
  \centering
  \captionsetup{justification=centering}
  \caption{TNDM architecture}
  \label{fig:tndm_arch}
\end{subfigure}
\caption{(a) The latent space of TNDM is partitioned into irrelevant and relevant high-dimensional initial conditions ${g_0}_i$ and ${g_0}_r$. These initial conditions are deterministically transformed to recover the latent factors $z_i$ and $z_r$ which give rise to the jointly observed neural activity $x$ and behaviour $y$. We assume there are K trials in the dataset. (b) TNDM utilizes a sequential variational autoencoding approach to ammortize inference of the relevant and irrelevant initial conditions ${g_r}_0$ and ${g_i}_0$. The initial conditions are passed through two separate RNNs to generate the behaviourally relevant and irrelevant dynamics $g_i$ and $g_r$ which are then projected into a low-dimensional subspace to recover the dynamical factors $z_i$ and $z_r$. These factors are used to reconstruct the neural activity and behaviour. The behaviour is reconstructed from the relevant factors using a flexible linear decoder which can capture complex temporal relationships (see the paragraph on \textbf{Behaviour Decoding}).}
\label{fig:figure1}
\end{figure}
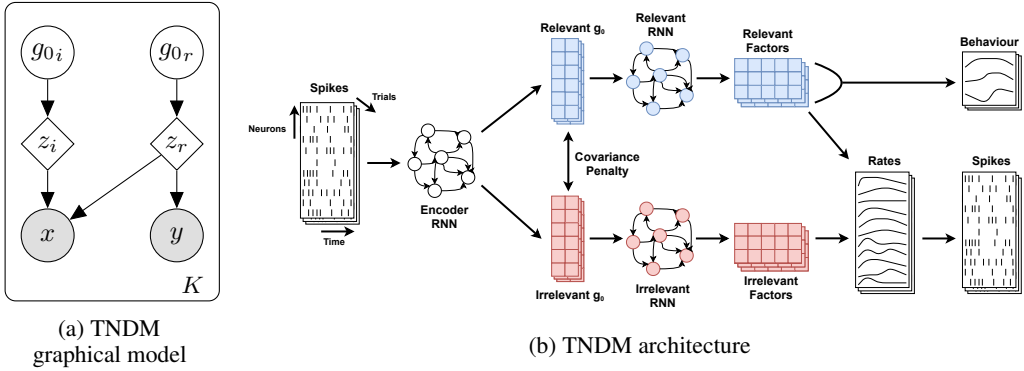

In this work, we introduce Targeted Neural Dynamical Modeling (TNDM). TNDM is a nonlinear state-space model which jointly models neural activity and an observed behaviour. Crucially, TNDM learns to reconstruct both the population activity and behaviour by disentangling the behaviourally relevant and behaviourally irrelevant dynamics that underlie the neural activity.

\paragraph{Generative model.} A plate diagram of TNDM's graphical model is shown in Figure \ref{fig:tndm_prob}. We assume that the observed neural activity $x$ and behaviour $y$ in each trial are generated by two sets of latent factors $z_i$ and $z_r$.

The generative process of TNDM is defined below:

\begin{equation}\label{eq:obs_model}
\begin{aligned}
&{g_i}_0 \sim p_{{g_i}_0}, {g_r}_0 \sim p_{{g_r}_0},\;
{g_i}_t = {f_\theta}_i({g_i}_{t-1}),\;
{g_r}_t = {f_\theta}_r({g_r}_{t-1})\\
&{z_i}_t = {W_i}_z({g_i}_t),\; {z_r}_t = {W_r}_z({g_r}_t),\; r_t = \exp(W_r({z_i}_t, {z_r}_t)))\\
&x_t \sim \text{Poisson}(x_t | r_t),\; y \sim \mathcal{N}(y | C_y(z_r), I)
\end{aligned}
\end{equation}

In the above equation, $p_{{g_i}_0}$ and $p_{{g_r}_0}$ are the distributions over the initial conditions of the behaviourally irrelevant and behaviourally relevant dynamics, respectively (assumed to be Gaussian). Similarly to LFADS, we parameterize the nonlinear transition functions ${f_\theta}_i$ and ${f_\theta}_r$ using RNNs\footnote{To implement TNDM, we primarily adapt the original Tensorflow \cite{tensorflow2015-whitepaper} implementation of LFADS from \url{https://github.com/lfads/models} (Apache License 2.0).}. As ${g_i}_t$ and ${g_r}_t$ can have arbitrarily high-dimensionality in our model (defined by the number of units in the RNN), we utilize two weight matrices, ${W_i}_z$ and ${W_r}_z$, to project these high-dimensional dynamics into a low-dimensional subspace, giving rise to the relevant and irrelevant dynamical factors ${z_i}_t$ and ${z_r}_t$ at each time step. These factors are then used to reconstruct both the observed neural activity and behaviour using linear decoders. An essential feature of our generative model is that although neural activity is reconstructed from the latent dynamics at each time step (i.e. no time lag), we let the relevant factors reconstruct the behaviour with a more flexible linear decoder $C_y$ that allows for time lags (explained in the \textbf{Behaviour Decoding} paragraph).

It is important to understand that although the dimensionality of the dynamics in TNDM (and LFADS) can be arbitrarily high, the dimensionality of the subspace that gives rise to neural activity and behaviour will be low due to the projection. Therefore, our model can be used to examine the number of latent variables (i.e. activity patterns) that are needed to characterize the population response and corresponding behaviour. As this is the primary goal when fitting LVMs to neural data \cite{cunningham2014dimensionality}, we compare all LVMs in this paper (TNDM, LFADS, and PSID) by the dimensionality of this subspace rather than the dimensionality of the dynamics.


\paragraph{Behaviour Decoding.} As mentioned above, PSID utilizes a linear weight matrix that maps the relevant latent dynamics at each time step to the behaviour at each time step, i.e $y_t = W_y(z_{r_t})$. This parameterization does not allow for modeling any latency, long-term dependencies or correlations, therefore, it severely limits the ability of the relevant dynamics to simultaneously explain neural activity and behaviour. To demonstrate the drawbacks of the no time lag behaviour decoder, we show that while training TNDM using this decoder leads to accurate behaviour prediction, the reconstruction of the neural activity noticeably decreases. This issue gets exacerbated in models with nonlinear dynamics as the expressivity of the underlying RNNs, along with the inflexible one-to-one behaviour mapping, leads the relevant dynamics to simply learn to replicate the behaviour of interest. These results are summarized in Supplement 1.

To overcome this limitation we instead allow the relevant latent dynamics to reconstruct the behaviour through any learned linear causal relationship. To this end, we introduce a linear weight matrix $C_y$ with dimensionality $n_{z_r}T\times BT$ where $n_{z_r}$ is the number of relevant dynamics, $B$ is the number of behaviour dimensions, and $T$ is the time length of a single trial. To transform the relevant factors $z_r$ into behaviours using $C_y$, we concatenate each dimension of $z_r$ in time to form a 1D vector $Z_r$ with length $n_{z_r}T$ and then perform the operation $Y = C_yZ_r$ where Y is the resulting concatenated behaviour. As $Y$ is a 1D vector with length $BT$, we can reshape $Y$ to recover the reconstructed behaviour $\hat{y}$. Importantly, we do not allow acausal connections in $C_y$, i.e. the lower triangular components of each of the dynamics to behaviour blocks are set to zero during training. For an example weight matrix $C_y$, see Figure \ref{fig:data3}a. In comparison to a simple no time lag mapping, we find that our flexible, causal linear decoder allows the relevant latent dynamics to both reconstruct the measured behaviour and capture neural variability. This is shown in Supplement 1 where the behaviourally relevant factors learned by TNDM with the full causal decoder contribute more meaningfully to the neural reconstruction than when using the no time lag decoder.

\paragraph{Inference}
To extract the latent dynamics $z_r$ and $z_i$ from the neural activity $x$, we learn to approximate the posterior over the initial conditions of the dynamics ${g_r}_0$ and ${g_i}_0$ and then we learn the RNN mapping from the initial conditions to the latent dynamics as together they deterministically define $z_r$ and $z_i$. To approximate the true posterior $p({g_r}_0,{g_i}_0| x, y)$, we implement TNDM as a sequential VAE and define our variational posterior as the product of two independent multivariate Gaussian distributions with diagonal covariances. The variational parameters of each Gaussians are computed with a shared encoder network $e_{\phi_0}$ followed by separate linear transformations:

\begin{align}
{q_\Phi}_r({g_r}_0 | x){q_\Phi}_i({g_i}_0 | x) = \mathcal{N}(\mu_{\phi_{r_1}}(e_{\phi_0}(x)), {\sigma^2}_{\phi_{r_2}}(e_{\phi_0}(x))) \cdot \mathcal{N}(\mu_{\phi_{i_1}}(e_{\phi_0}(x)), {\sigma^2}_{\phi_{i_2}}(e_{\phi_0}(x)))
\end{align}

The inference networks for the behaviourally relevant and the behaviourally irrelevant initial conditions are parameterized by $\Phi_r =\{\phi_0,\phi_{r_1},\phi_{r_2}\}$ and $\Phi_i =\{\phi_0,\ \phi_{i_1},\phi_{i_2}\}$, respectively. It is important to note that TNDM's variational posterior only depends on the neural activity $x$. This approximation forces the learned initial conditions to come from the observed activity and allows for decoding of unseen behaviours after training. The reparameterization trick is used to sample from each initial condition distribution and the sampled initial conditions are evolved using separate decoder RNNs to produce the behaviorally relevant ${g_r}_t$ and irrelevant high-dimensional dynamics ${g_i}_t$. The high-dimensional dynamics at each time-step are projected into a low-dimensional subspace to recover the low-dimensional dynamical factors ${z_r}_t$ and ${z_i}_t$. The neural activity $x$ and behaviour $y$ are generated from the latent factors $z$ as shown in Equation \ref{eq:obs_model}.

The ELBO for the observed data from a single trial is therefore defined as:

\begin{align}\label{eq:tndm_elbo}
\text{ELBO}(x,y) = - \KL{{q_\Phi}_r}{p_{{g_r}_0}} - \KL{{q_\Phi}_i}{p_{{g_i}_0}} + \mathbb{E}_{{q_\Phi}_r{q_\Phi}_i} [\log p_{\theta_1}(x|g_i, g_r)p_{\theta_2}(y|g_r)]
\end{align}

where $p_{\theta_1}$ and $p_{\theta_2}$ are the observation models for the neural activity and behaviour, respectively.

\paragraph{Disentangling the latent dynamics} Despite factorising the variational posterior, the true posterior over the latent variables $P({g_0}_i, {g_0}_r | x, y)$ cannot be factorized; that is, $z_i$ and $z_r$ (which are deterministic transforms of ${g_0}_i$ and ${g_0}_r$) are conditionally dependent given the observed data. This means that $z_i$ and $z_r$ will be statistically dependent. To reduce sharing of information and redundancy between these two sets of dynamics, we introduce a novel disentanglement penalty on the two initial condition distributions. For this penalty, we take inspiration from the domain of unsupervised disentangled representation learning where it is standard to introduce additional penalties that encourage disentanglement or independence of the latent representations \cite{kumar2017variational, chen2018isolating,kim2018disentangling}. As mutual information
is hard to compute for high-dimensional variables and the experimental data often has a very limited number of trials to estimate these distributions reliably, we instead penalize the mean of the sample cross-correlations between ${g}_{0_r}$ and ${g}_{0_i}$. Importantly, this cross-correlation penalty is applied in such a way that the final objective is still a valid lower bound of the log-likelihood (the cross-correlation penalty is always negative).
We refer to this penalty as $Q({q_\Phi}_r, {q_\Phi}_i)$ and we adjust its weight with a hyperparameter $\lambda_{Q}$ (see Supplement 4 for an ablation study of this penalty). The final objective function for TNDM is then:

\begin{align}\label{eq:tndm_objective}
J(x,y) = &-\KL{{q_\Phi}_r}{p_{{g_r}_0}} - \KL{{q_\Phi}_i}{p_{{g_i}_0}} + \mathbb{E}_{{q_\Phi}_r{q_\Phi}_i} [\log p_{\theta_1}(x|g_i, g_r)]\nonumber + \\ &\lambda_b\mathbb{E}_{{q_\Phi}_r}[\log p_{\theta_2}(y|g_r)] + \lambda_{Q}Q({q_\Phi}_r, {q_\Phi}_i)
\end{align}
where $\lambda_b$ is an additional hyperparameter introduced to balance the behavioural likelihood with the neural likelihood (see Supplement 3 for hyperparameter details). While TNDM is not the first VAE to jointly model two observed variables with a partitioned latent space \cite{whiteway2021partitioning}, it is distinguished by its unique objective function and penalties, its RNN-based architecture, its causal linear decoder, and its novel application to neural activity and behaviour.

\section{Experiments}
\label{sec:experiments}
\begin{figure}
\centering
  \includegraphics[scale=.65]{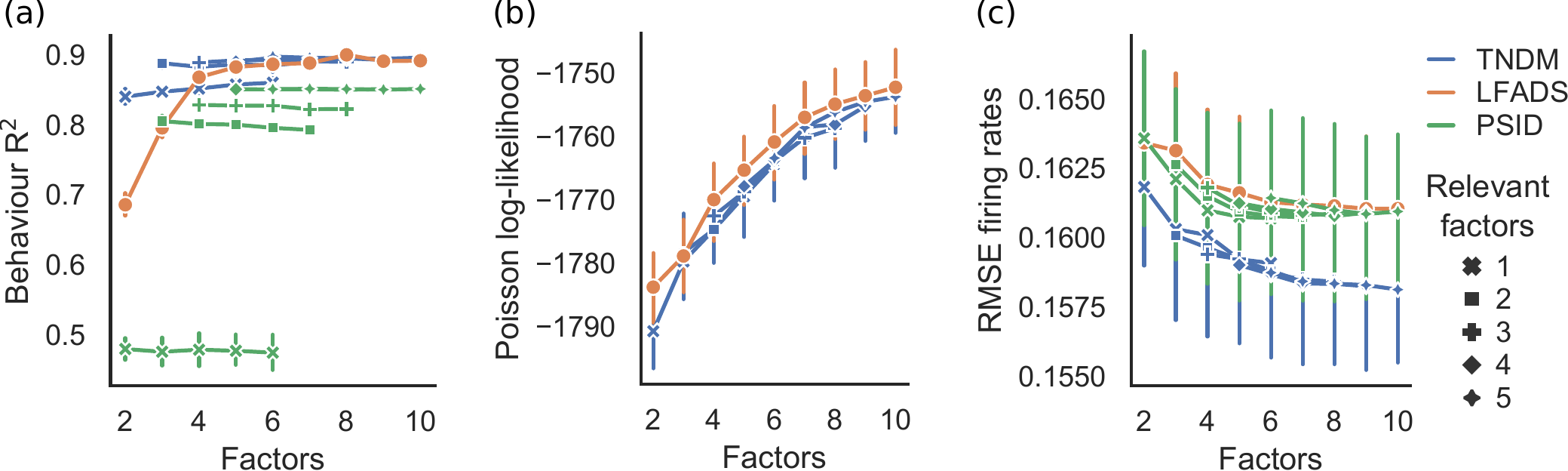}
\caption{Summary of the behaviour and activity reconstruction accuracy for TNDM, LFADS, and PSID fit to neural recordings from the monkey primary motor cortex (M1) during a center-out reaching task. Each plot shows performance as a function of the total number of latent factors, averaged over five fits with different initialisation (random seeds) and different random training-test data splits. Error bars show standard error of the mean. (a) Coefficient of determination (R$^2$) for measured and reconstructed behaviour (hand position). For TNDM, the reconstruction performed by the behaviour decoder (with only the relevant factors) while for LFADS, a ridge regression had to be used to decode the behaviour ex post factor. For PSID the reconstruction by the model was additionally Kalman smoothed. (b) Poisson log-likelihood for the activity reconstruction per single trial for TNDM and LFADS. (c) Root mean square error (RMSE) between the predicted and actual ground-truth firing rates. Averaging was performed across all trials with the same movement direction. Behaviour reconstruction and log-likelihood were computed on held out test data, and the firing rate RMSE on the whole data set to allow for more reliable averaging.}
\label{fig:data1}
\end{figure}

\subsection{Simulated Data}

We evaluate TNDM on synthetic spike trains generated from a Lorenz system, a common benchmark for state space models of neural activity. For a detailed description of the simulated data and evaluation, we refer the reader to Appendix 6.

\subsection{M1 neural recordings during reach}

We apply TNDM to data gathered from a previously published monkey reaching experiment \cite{gallego2020long}. The monkey is trained to perform a center-out reaching task with eight outer targets. On a go cue, the monkey moves a manipulandum along a 2D plane to the presented target and receives a liquid reward upon success. Spiking activity from M1 and PMd along with the 2D hand position are recorded during each trial. We train each model on single-session data gathered from one of the six trained monkeys. The data consist of two paired datasets: PMd activity paired with hand position and M1 activity paired with hand position. We show results for the M1 recordings in the main text and the results for the PMd recordings in Supplement 2. The neural activity is counted in 10ms bins and the behaviour is also measured every 10ms. We align the behaviour to the spikes for both datasets by taking the activity starting during movement onset. We set the length of the neural activity to be the minimum time until the target is reached across all trials.  As one of our baselines, PSID cannot model spike count data, we smooth the spike counts with a Gaussian kernel smoother (with standard deviation 50ms) before applying PSID. Out of the 176 trials from the experiment, we use 80\% for training (136 trials). We hold out the remaining 34 trials to test the models.

\paragraph{Models/Evaluation} For all models, we perform a sweep over the number latent factors. For TNDM and PSID, we train models with all combinations of 1-5 relevant latent factors and 1-5 irrelevant factors (e.g. 3 relevant and 2 irrelevant). For LFADS, we train models with the number of latent factors ranging from 2-10. As TNDM and LFADS are both implemented as sequential variational autoencoders, we fix the architectures to be same for the two methods (64 units in the generators and encoder). We fix all shared hyperparameters to be the same between the two methods except for the dropout (TNDM requires more regularization due to the behavioural loss). For a detailed explanation of the hyperparameters and architectures used in these experiments, see Supplement 3. All results reported here are based on five fits of each model with different random seeds and data splits.

To compare TNDM to LFADS, we first evaluate their neural reconstruction using the test data Poisson log-likelihood and the root mean square error (RMSE) between the predicted and actual ground-truth firing rates. To calculate the ground-truth firing rates, we average the neural data across all trials with the same movement direction and used both the training and test sets to get more robust estimates of the rates from the experimental data. To evaluate the behaviour reconstruction of LFADS, we perform an ex post facto regression from the extracted latent factors to the behaviour in the training set. This regression is linear and is from all time steps of the factors to all time steps of the behaviour\footnote{We utilize a standard ridge regression from scikit-learn \cite{scikit-learn} with default parameters.}. Note that this approach for regressing the LFADS factors is more flexible than the decoder in TNDM which is also linear but constrained to be causal. We then compute the coefficient of determination (R$^2$) between the decoded and ground-truth behaviour for each model on the test data.

To compare TNDM to PSID, we evaluate the neural reconstruction by computing the RMSE between the predicted and actual ground-truth firing rates. For behaviour reconstruction, we compute the R$^2$ between the decoded and ground-truth behaviours for each model on the test data.\footnote{A potential concern when comparing TNDM and PSID on behaviour reconstruction is that TNDM has more parameters in its behaviour decoder than PSID does. This is because TNDM decodes the behavioural variable at time step $t$ using all past time steps of the latent factors while PSID only uses the current time step $t$. As shown in Supplement 1, however, TNDM achieves equally high behaviour reconstruction using the no time lag decoder as it does using the proposed casual decoder, therefore, the number of parameters in TNDM's behaviour decoder is not a confounding factor for this evaluation. Also, while it is possible to remap PSID's learned latent factors to the behaviour using a higher parameter regression, this would be equivalent to changing PSID's generative model and, therefore, would no longer be a valid comparison to PSID.} As both TNDM and LFADS use information from the whole trial to infer the latent factors (which is inherently acausal), we use a standard Kalman smoother for PSID to make the state estimation comparable. 

\section{Results}
\label{sec:results}

\subsection{Simulated Data}

For a detailed discussion of TNDM's results on synthetic spike trains generated from a Lorenz system, we refer the reader to Appendix 6.

\subsection{M1 neural recordings during reach}

\paragraph{Fit to behaviour}
The behaviour reconstruction of TNDM, LFADS and PSID is summarized in Figure \ref{fig:data1}a. LFADS behavioural reconstruction saturates at around eight factors ($R^2\approx0.89$) and with just four factors yields a respectable behavioural fit ($R^2\approx0.86$). This indicates that the LFADS factors, which are constrained only by neural activity, are interpretable in terms of externally measured variables. In comparison, TNDM achieves saturating performance with just three latent factors ($R^2\approx0.90$) where \textit{only two are constrained to be relevant} for behavioural reconstruction. In fact, all TNDM models with three or more factors (where at least two are constrained to be relevant) have similar behaviour reconstruction accuracy. In comparison to TNDM, LFADS achieves a behaviour reconstruction of just ($R^2\approx0.68$) for three latent factors. TNDM also has much more accurate behaviour reconstruction on the test data than PSID. For three latent factors, where two are constrained to be relevant, PSID achieves a behavioural fit of ($R^2\approx0.82$). PSID's behavioural reconstruction saturates at six latent factors where five are constrained to be relevant $R^2\approx0.88$). Overall, TNDM's behaviour decoding performance implies that the dimensionality of the behaviourally relevant dynamics for this 2D task are lower-dimensional than previously predicted by other latent variable modeling approaches.



\paragraph{Fit to neural activity} Do the additional constraints and penalties of TNDM affect the accuracy of neural activity reconstruction? This is an important question to answer as the learned latent dynamics are only meaningful if they also explain the observed activity well. 
Surprisingly, we find that TNDM's and LFAD's Poisson log-likelihoods on the test data are very close (Figure \ref{fig:data1}b). This indicates that the partitioning of the latent dynamics and the additional constraints imposed by TNDM have a very small effect on its neural reconstruction for this dataset. Instead, TNDM and LFADS both show a gradual improvement of neural activity reconstruction as a function of the number of factors. The only deviation from this trend is TNDM with one relevant and one irrelevant factor. This is not surprising, however, as much of the neural variability is explained by the behaviour of interest; only allowing one latent factor to explain the neural variability related to behaviour (while simultaneously enforcing disentanglement between the relevant and irrelevant dynamics) will cause TNDM's neural activity reconstruction to suffer. Perhaps more surprisingly, TNDM achieves a lower firing rate RMSE than LFADS with the same number of factors (Figure \ref{fig:data1}c). While on the surface this result seems counterintuitive, it may be because the RMSE is computed for the average firing rate over all trials of the same movement direction. While TNDM and LFADS have a very similar Poisson likelihood on single trials, TNDM can better distinguish trials by movement direction since it is explicitly modeling behaviour, hence the firing rate prediction split by trial type is improved. PSID provides a worse fit to the neural data than TNDM which is expected given that it is constrained to learn linear dynamics. 

\paragraph{PSID failure mode} Although the neural reconstruction is fairly good for PSID, we find an unexpected result when analyzing PSID's learned model parameters. Specifically, we find that PSID's state-transition matrix $A$, which characterizes the behaviorally relevant neural dynamics \cite{saniModelingBehaviorallyRelevant2020}, is approximately the identity matrix for this dataset and is non-informative about the neural activity. We expand upon this analysis of PSID in Supplement 5 where we show that PSID recovers the same state-transition matrix $A$ when we shuffle the neural data by trial or by time. We provide further evidence that PSID is unable to find informative linear dynamics for this dataset because the behaviour is inherently nonlinear across trials (i.e. multi-directional reaches). Therefore, on this dataset, we conclude that PSID's performance on neural reconstruction is mainly determined by the behaviourally irrelevant factors and its performance on behaviour reconstruction is completely determined by the Kalman gain during decoding.


\begin{figure}[t]
\centering
  \includegraphics[scale=.75]{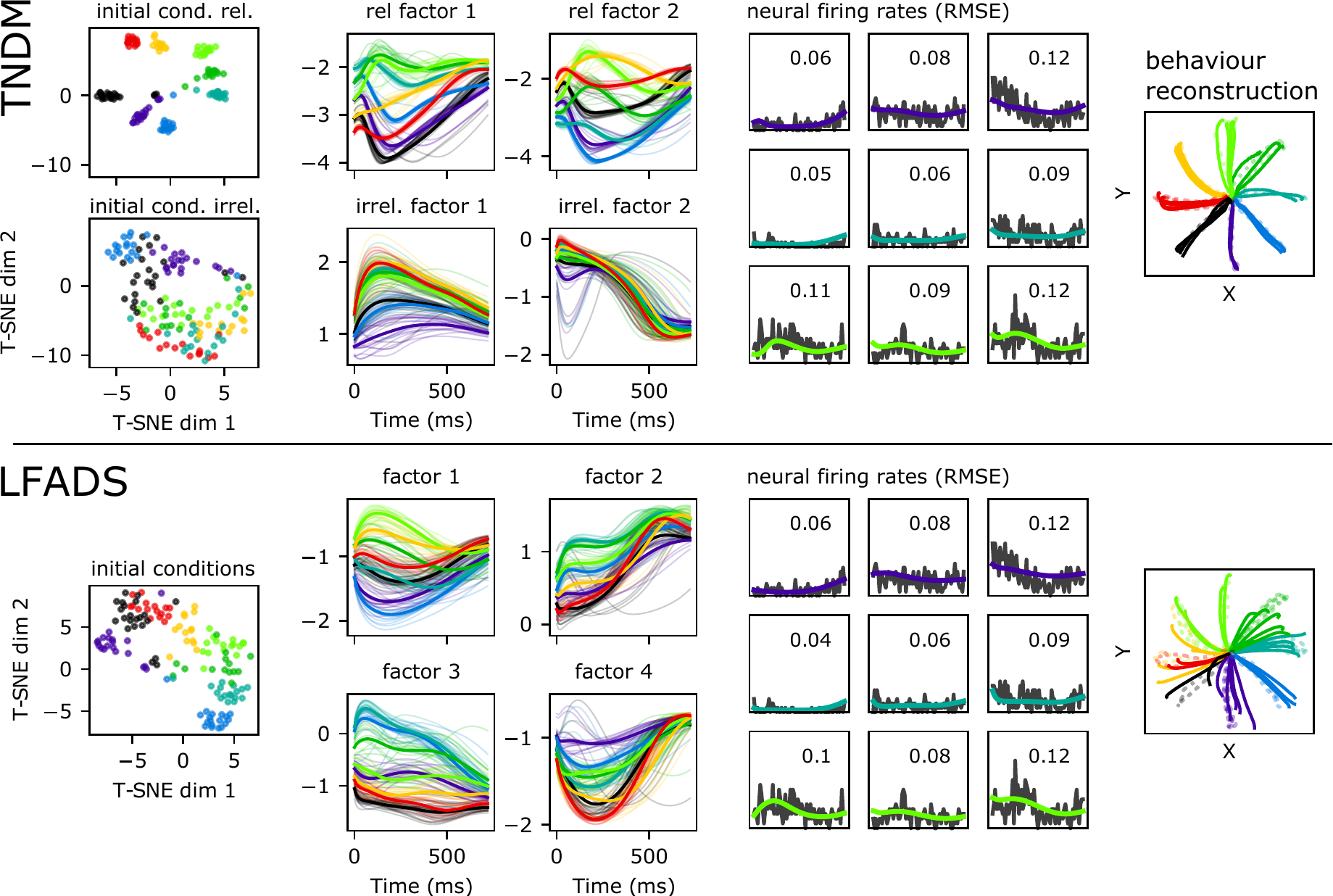}
\caption{We visualize each component of the generative process for TNDM (top) and LFADS (bottom) after training. On the far left, we visualize the inferred initial conditions for each method after reducing the dimension to 2 with T-SNE. As can be seen, TNDM's inferred initial conditions show a clear distinction between behaviourally relevant and behaviourally irrelevant information whereas LFADS inferred initial conditions mix this information together. Next, we show the condition-averaged inferred latent dynamical factors (along with the single-trial factors) for each method to demonstrate that there is a clear distinction between the behaviourally relevant and behaviourally irrelevant factors in TNDM but not in LFADS. Finally, we show neural activity reconstruction (numbers are RMSE between data and prediction) and behaviour reconstructions (linear regression for LFADS) for both methods to illustrate that TNDM provides an excellent fit to the neural data despite the partitioned latent space and behavioural prediction.}
\label{fig:data2}
\end{figure}



\begin{figure}
\centering
  \includegraphics[scale=.7]{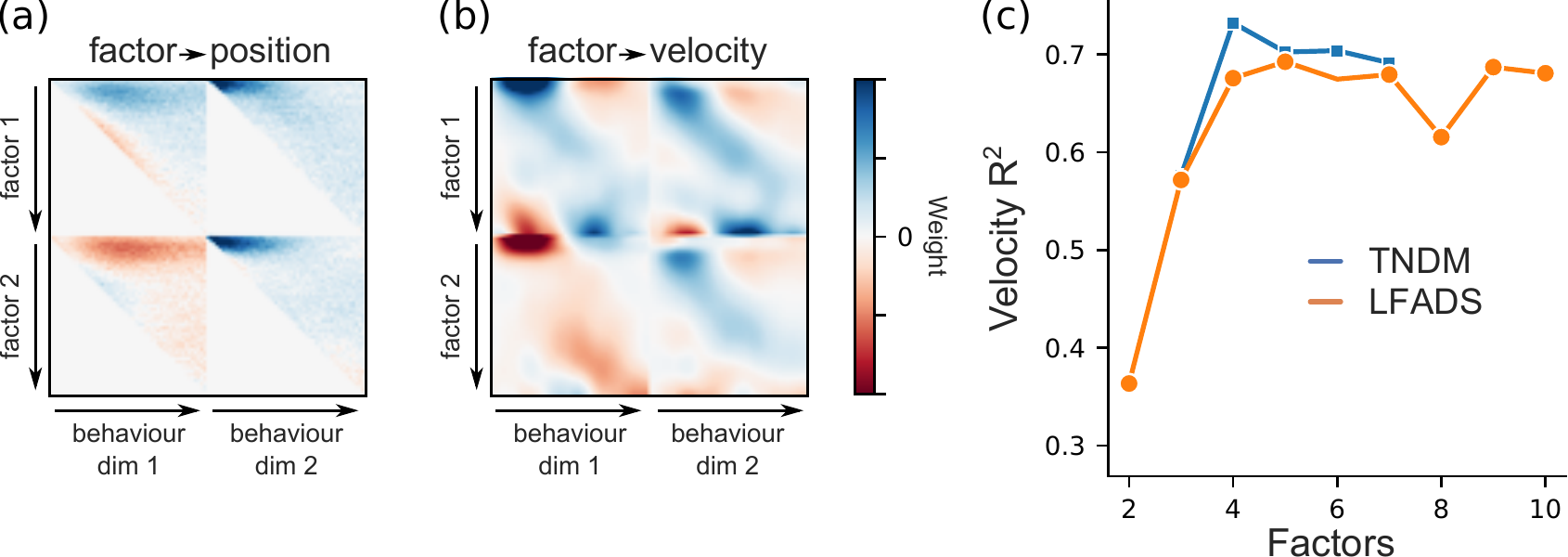}
\caption{(a) Visualization of the weights of the TNDM behaviour decoder $C_y$ that transforms two relevant latent factors into behaviour (hand position). The behaviours are aligned horizontally and the factors vertically. The upper-triangular structure reflects causal decoding, i.e. factors can only influence future behaviour. The model had two relevant and two irrelevant factors. (b) Weights obtained by using ridge regression to predict movement velocity (x and y components) from the relevant factors. A diagonally banded structure can be observed indicating a (delayed) identity transformation. Unlike in (a), this matrix is not constrained to be causal. (c) Decoding accuracy for velocity obtained using ridge regression for LFADS and TNDM with two relevant factors and a varying number of irrelevant factors.}
\label{fig:data3}
\end{figure}

\paragraph{Interpretation of the learned dynamics} In Figure \ref{fig:data2}, we visualize each stage of the generative process of TNDM (2 relevant and 2 irrelevant factors) and LFADS (4 factors) after training both models on the M1 dataset. As can be seen in the figure, there appears to be a clear separation between the relevant and irrelevant initial condition distributions in TNDM that is less apparent in the mixed latent space of LFADS. In fact, the relevant initial conditions of TNDM seem to precisely capture the reach direction of the behaviour. Despite the noticeable differences between the dynamics of LFADS and TNDM, their ability to infer the underlying firing rates from this dataset are nearly identical.

Looking at the learned dynamical factors for TNDM, one can see that the relevant dynamics are more clearly distinguished by reach condition and there is much less variance in the learned trajectories than those of LFADS. At the same time, the relevant TNDM factors do not trivially re-capitulate the behaviour dynamics, indicating that the dual constraint of behaviour and neural activity unmasks a more complicated relationship between the two. This relationship can be analysed by visualizing the learned weights of TNDM's behaviour decoder as shown in Figure \ref{fig:data3}a (for two relevant factors and two irrelevant factors). In this weight matrix, each time point of the behaviour receives contributions from a broad time interval of preceding factor activity. This corresponds to a temporal integration of the factors and suggests that the relevant factors represent information about movement velocity. Indeed, velocity can be decoded well from the relevant factors using a simple ridge regression (both for TNDM and LFADS, Figure \ref{fig:data3}c). The learned coefficients of this ridge regression for TNDM have a diagonally banded structure that corresponds to a delayed identity transformation (Figure \ref{fig:data3}b), which is not visible for the LFADS factors (not illustrated). Taken together, these results suggest that M1 neural dynamics are related to velocity of the hand in this task. Interestingly, we find that velocity decoding peaks at two relevant factors for TNDM and is less discernible when this number is increased, indicating that the addition of more relevant factors may spread this velocity information across the factors in a nonlinear way which cannot be recovered by the ridge regression (not visualized in Figure \ref{fig:data3}). It also illustrates that the TNDM's behaviour prediction saturation point (two relevant factors) has perhaps the most interpretable latent space of all the trained TNDM models.



The irrelevant factors in TNDM show task-aligned dynamics that do not depend strongly on the task type, but are rather homogeneous (see Figure \ref{fig:data2}). For instance, over the course of each trial, irrelevant factor 2 has a large initial fluctuation followed by a steady increase in its absolute value over time until around 600ms where it tapers off (around when the monkey reaches the target destination). As this factor is agnostic to the reach direction, this may reflect dynamics associated with execution of these movements more generally. 




\section{Discussion}
\label{sec:discussion}
In this work, we introduce TNDM, a nonlinear state-space model designed to disentangle the behaviourally relevant and behaviourally irrelevant dynamics underlying neural activity. We evaluated TNDM on synthetic data and on neural population recordings from PMd and M1 paired with a 2D center-out reach behaviour. We showed that TNDM was able to extract low-dimensional latent dynamics that were much more predictive of behaviour than those of current state-of-the-art state-space models without sacrificing its fit to the neural activity. This led us to the interpretation that the dimensionality of the neural activity associated with the 2D reaching task is potentially lower than previously thought and may be associated with the velocity of the hand. 

Although the initial results presented for TNDM are quite promising, the method has a few limitations that should be addressed. First, we find that some hyperparameter settings combined with certain random initialisations can cause biologically implausible oscillations in the learned latent dynamics. While more work needs to be done to understand this, it could be related to the weighting between the behavioural and neural likelihoods or to the capacity of the model. A second limitation of TNDM is whether the disentanglement penalty between the relevant and irrelevant dynamics is sufficient. Although the covariance penalty works well in practice (on the presented datasets), disentangling sources of variation using deep generative models is still an open problem \cite{locatello2019challenging}. Similarly to PSID, TNDM could be implemented in multiple separate stages which may allow for better disentanglement of the relevant and irrelevant dynamics \cite{srivastava2020improving}. Third, the linear causal decoder we introduce for behaviour reconstruction is parameter inefficient: the number of parameters scales quadratically with time and dynamics/behaviour dimension. Lastly, it can be challenging to determine the 'correct' latent dimensionalities for TNDM. For the datasets in this paper, we found that performing a wide sweep over a number of relevant and irrelevant dimensionalities and then choosing the relevant dimensionality where the behaviour prediction saturates is a potential recipe for finding an interpretable latent space.

In future work, we plan to train TNDM with higher-dimensional behaviours such as joint angle or electromyography (EMG) recordings. We also plan to extend TNDM to non-temporal/discrete behaviours which are of interest in behavioural neuroscience (e.g. decision making). Finally, we hope to extend TNDM such that it can model dynamics with external inputs from another brain region, i.e. non-autonomous dynamics.

\section{Broader Impact}
\label{sec:impact}
In this work, we develop an expressive and interpretable model of neural population activity. As such, we imagine that TNDM will be useful for answering important questions about neural function (e.g. how the motor cortex gives rise to behaviour). We also imagine that the ability of TNDM to accurately model both the neural activity and the behaviour of interest will be of interest for the brain-computer interface community. We believe that TNDM (or the principles behind it) can be used to improve behaviour decoding from neural activity. We also hope that TNDM inspires more research into deep generative models of neural activity that incorporate in external variables of interest. A possible negative societal impact of TNDM is that, like all deep neural network models, it requires a relatively large amount of compute and has a noticeable carbon footprint. 

\newpage

\section*{Acknowledgements}
We thank Alessandro Facchin and Nina Kudryashova for the code contributions and for the insightful discussions. We also thank the reviewers for their thoughtful critiques and suggestions.

\bibliographystyle{rusnat}
\bibliography{neurips_2021}

\newpage

\section*{Checklist}

\begin{enumerate}

\item For all authors...
\begin{enumerate}
  \item Do the main claims made in the abstract and introduction accurately reflect the paper's contributions and scope?
    \answerYes{}
  \item Did you describe the limitations of your work?
    \answerYes{\bf{See Discussion section}}.
  \item Did you discuss any potential negative societal impacts of your work?
    \answerYes{\bf{See broader impact section}}.
  \item Have you read the ethics review guidelines and ensured that your paper conforms to them?
    \answerYes{}
\end{enumerate}

\item If you are including theoretical results...
\begin{enumerate}
  \item Did you state the full set of assumptions of all theoretical results?
    \answerNA{}
	\item Did you include complete proofs of all theoretical results?
    \answerNA{}
\end{enumerate}

\item If you ran experiments...
\begin{enumerate}
  \item Did you include the code, data, and instructions needed to reproduce the main experimental results (either in the supplemental material or as a URL)?
    \answerNo{\bf{The datasets we use are large and not publicly available yet. We plan to release our codebase upon acceptance}.}
  \item Did you specify all the training details (e.g., data splits, hyperparameters, how they were chosen)?
    \answerYes{\bf{In the supplement and in the main text we specify parameters and splits.}}
	\item Did you report error bars (e.g., with respect to the random seed after running experiments multiple times)?
    \answerYes{\bf{We provide standard deviations for our results in the supplementary tables}}.
	\item Did you include the total amount of compute and the type of resources used (e.g., type of GPUs, internal cluster, or cloud provider)?
    \answerNo{\bf{The models we train are relatively small and compute should not be an issue.}}
\end{enumerate}

\item If you are using existing assets (e.g., code, data, models) or curating/releasing new assets...
\begin{enumerate}
  \item If your work uses existing assets, did you cite the creators?
    \answerYes{\bf{See footnote 3}}
  \item Did you mention the license of the assets?
    \answerYes{\bf{See footnote 3}}
  \item Did you include any new assets either in the supplemental material or as a URL?
    \answerNo{}
  \item Did you discuss whether and how consent was obtained from people whose data you're using/curating?
    \answerNA{}
  \item Did you discuss whether the data you are using/curating contains personally identifiable information or offensive content?
    \answerNA{}
\end{enumerate}

\item If you used crowdsourcing or conducted research with human subjects...
\begin{enumerate}
  \item Did you include the full text of instructions given to participants and screenshots, if applicable?
    \answerNA{}
  \item Did you describe any potential participant risks, with links to Institutional Review Board (IRB) approvals, if applicable?
    \answerNA{}
  \item Did you include the estimated hourly wage paid to participants and the total amount spent on participant compensation?
    \answerNA{}
\end{enumerate}

\end{enumerate}

\clearpage
\section*{Appendix}

\appendix
\renewcommand{\thesection}{\arabic{section}}

\section{Behaviour decoder}
\begin{figure}[h]
\centering
  \includegraphics[scale=.35]{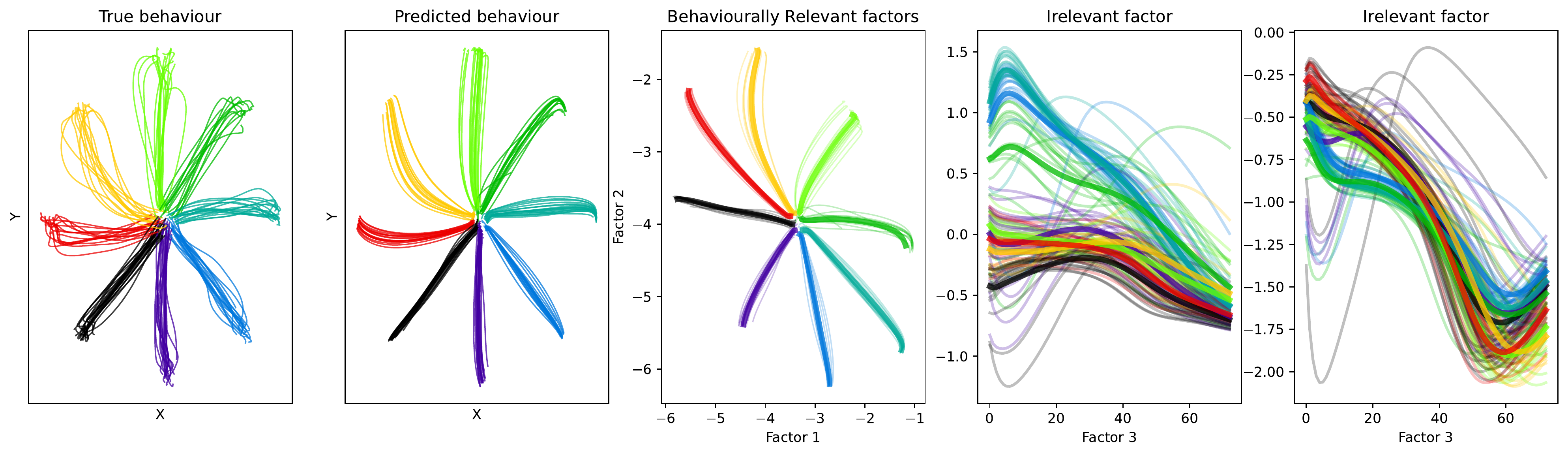}
  \includegraphics[scale=.35]{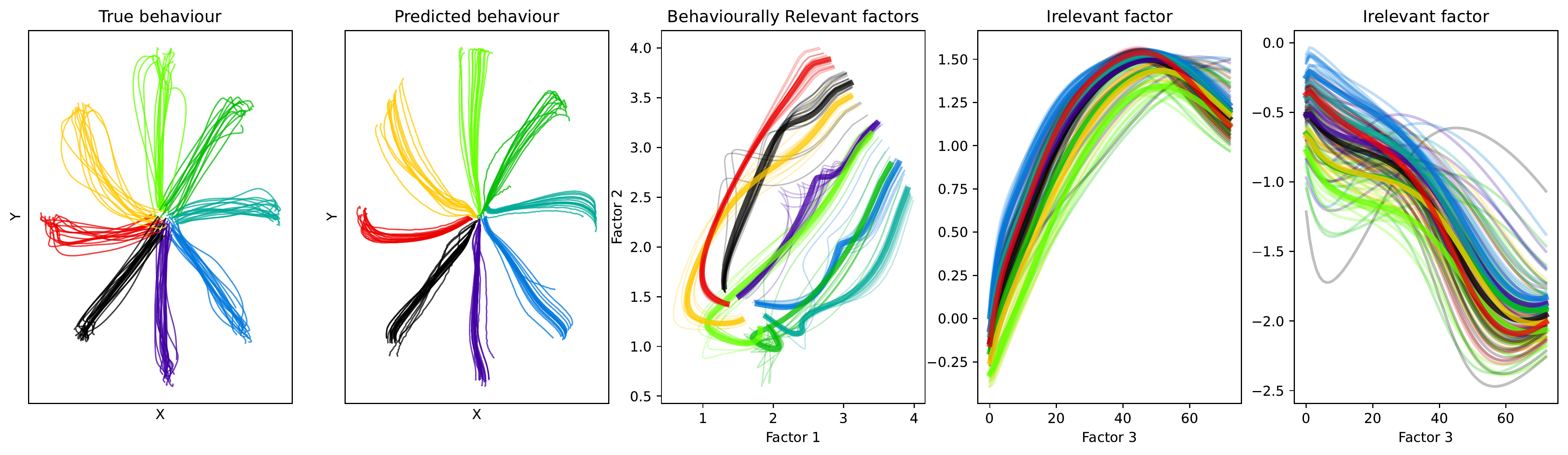}
\caption{Visualizations of the predicted behaviour and latent factors for TNDM trained with two different decoders. On top, TNDM is trained with a one-to-one Gaussian decoder (introduced in Section 2). As can be seen, the lack of flexibility in the decoder forces the relevant factors to simply replicate the behaviour. This means that the irrelevant factors have to encode behavioural information since they are primarily used for neural reconstruction. On bottom is the causal linear decoder introduced in this work. Here, the relevant factors capture more structure in the neural activity while still allowing for good behaviour reconstruction. This lets the irrelevant factors encode less behavioural information.}
\label{fig:decoders}
\end{figure}

\begin{figure}[h]
\centering
  \includegraphics[scale=.5]{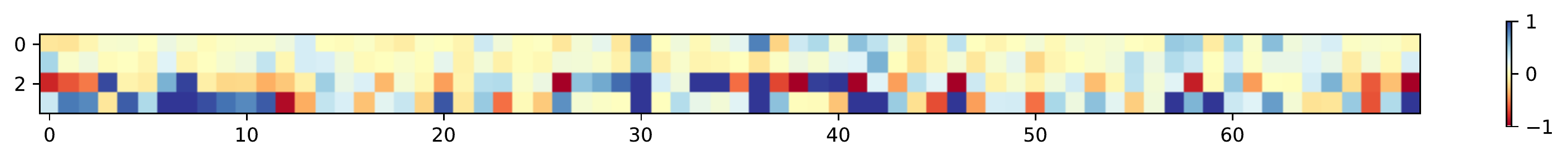}
  \includegraphics[scale=.5]{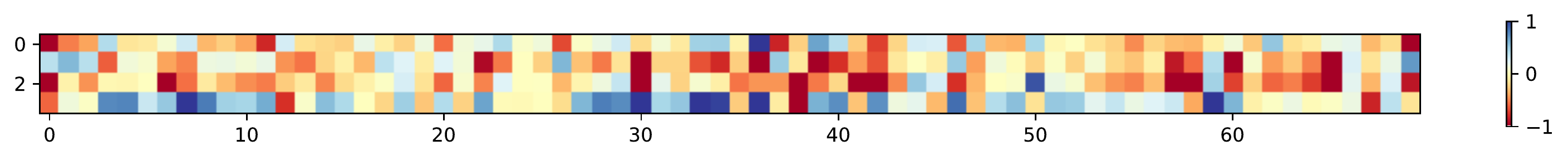}
\caption{Visualization of the learned neural reconstruction weight matrices for the factors shown in Figure \ref{fig:decoders}. These matrices transform the learned factors into neural firing rates. On top is the neural weight matrix for TNDM trained with a one-to-one Gaussian decoder and on bottom is the neural weight matrix for TNDM trained with the linear causal decoder. In both cases, their are four factors that are transformed into the firing rates of 70 neurons (the top two factors are relevant and the bottom two are irrelevant). Interestingly, for TNDM trained with the one-to-one decoder, the relevant factors are barely used for neural reconstruction (i.e. low magnitude weights). This implies that the learned factors are not informative of neural activity and that the irrelevant factors are mainly being used. For TNDM trained with the linear causal decoder, however, the relevant factors play a much larger role in the neural reconstruction (i.e. higher magnitude weights).}
\label{fig:neural_recon_decoders}
\end{figure}

As mentioned in Section 3, the choice of behaviour decoder can dramatically change the learned relevant latents. Figure \ref{fig:decoders} shows TNDM trained with both the one-to-one Gaussian behaviour decoder introduced in PSID and with the flexible, casual decoder introduced in this work. As can be seen with the one-to-one Gaussian decoder, TNDM's behaviourally relevant factors perfectly replicate the behaviour. This is because there is so little flexibility for the decoder to model the behaviour that replicating the behaviour is the best option (despite the potential negative effect to neural reconstruction). This forces the irrelevant factors to encode behavioural information as the neural reconstruction will largely depend on the irrelevant factors (see Figure \ref{fig:neural_recon_decoders}). With the causal linear decoder, TNDM is able to extract behaviourally relevant factors that explain the behaviour while still contributing meaningfully to the neural reconstruction (see Figure 
\ref{fig:neural_recon_decoders}). The irrelevant factors, in this case, do not need to encode behavioural information and can instead encode aspects of the neural activity unrelated to the specific behaviour. We find that with only one irrelevant factor, the neural reconstruction suffers for the one-to-one decoder and improves steadily as you add more irrelevant factors.

\clearpage
\section{Premotor cortex results}

\begin{figure}[h]
\centering 
  \includegraphics[scale=.7]{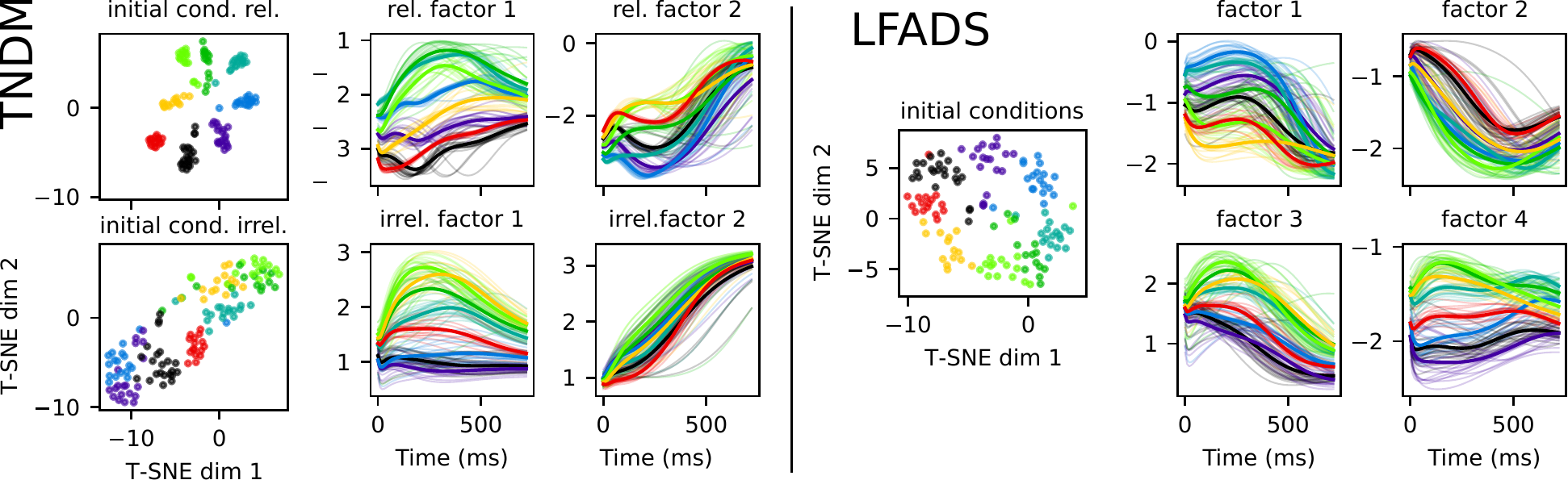}
\caption{Visualization of initial conditions and latent dynamical factors for a model of premotor cortex (PMd) activity. The activity was recorded simultaneously with the motor cortex (M1) activity shown in Figure \ref{fig:data2}, main text. As in that figure, TNDM (left) has two relevant and two irrelevant factors, and LFADS (right) four factors. On the left inferred initial conditions for each method are shown after reducing the dimension to 2 with T-SNE. There is a clear distinction between the conditions relating to different movement directions (indicated by different colours) in the relevant factors, but unlike for M1 the irrelevant factors also contain some structure that distinguish trial types. The LFADS initial conditions for PMd show some distinction between behaviours, but similar to the M1 data this is much weaker than for the relevant factors in TNDM. Condition-averaged inferred latent dynamical factors (along with the single-trial factors; plots on the right) again show a clear distinction between different behaviours for the relevant factors, while this information is mixed in the irrelevant factors. As for M1, there is no clear behaviour separation in the factors of LFADS.}
\label{fig:pmd_examples}
\end{figure}

\begin{figure}[h]
\centering 
  \includegraphics[scale=.65]{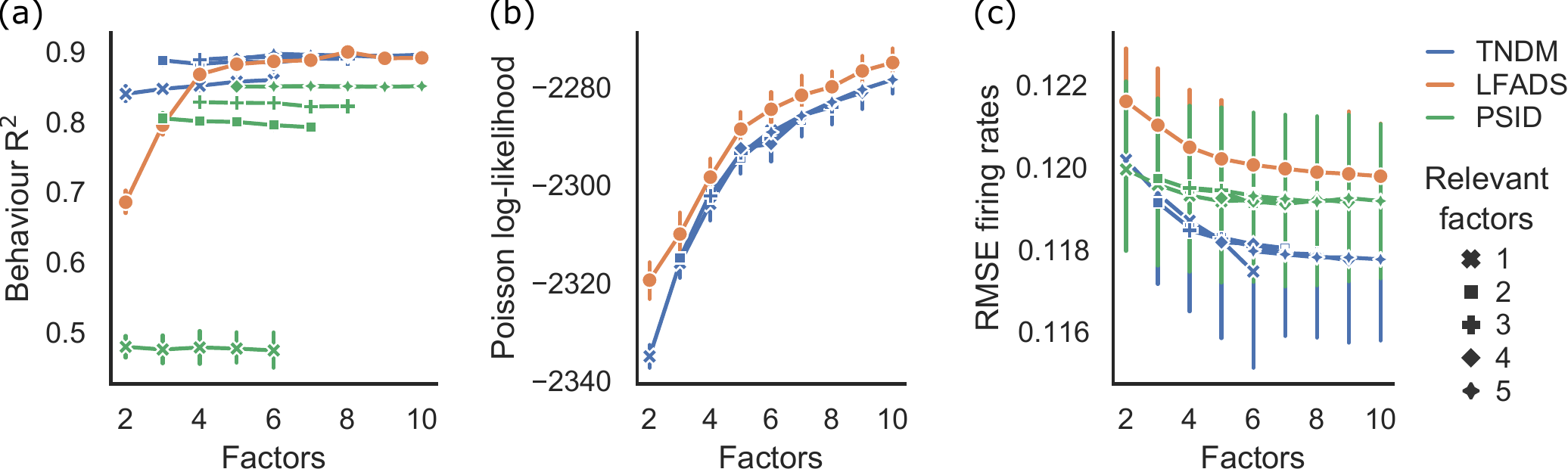}
\caption{Behaviour and activity reconstruction accuracy for TNDM, LFADS, and PSID for monkey premotor cortex (PMd) activity during the center-out reaching task. The data used here was recorded simultaneously with the M1 activity shown in Figure \ref{fig:data1} (main text). Each plot shows performance as a function of the total number of latent factors, averaged over five fits with different initialisation (random seeds) and different random training-test data splits. Error bars show standard error of the mean.
For TNDM and PSID, the reconstruction performed by the behaviour decoder (with only the relevant factors) is shown, while for LFADS a ridge regression was used to decode the behaviour ex post factor. Behaviour reconstruction and log-likelihood were computed on held out test data, and the firing rate RMSE on the whole data set to allow for more reliable averaging. Similar to M1, TNDM requires at least two relevant factors for saturating behaviour reconstruction accuracy for all model sizes. LFADS gradually reaches peak accuracy, saturating at 5 factors, and PSID requires at least five relevant factors. As in M1, neural activity reconstruction in TNDM solely depends on the total number of factors, irrespective of the fraction of relevant factors.}
\label{fig:pmd_all_results}
\end{figure}

\clearpage
\section{Hyperparameters}
Table \ref{table:hyperparams} shows the hyperparameters of LFADS and TNDM used for the main experiments. We did not run an exhaustive search over these parameters. For LFADS, we used default parameters for all regularization terms. For TNDM, we used a small value for $\lambda_b$ such that the behaviour likelihood was smaller than the neural likelihood (the neural reconstruction was the primary goal). The dropout for TNDM was set to be slightly higher than for LFADS so as to not overfit the behaviour; we found that this higher dropout was not helpful for LFADS. We also set the batchsize to 16 for TNDM and 10 for LFADS; we found that LFADS latent factors were less informative about behaviour when using a higher batch size. We qualitatively found that both models provided a good fit to the neural data with these parameters (see Figures \ref{fig:data2}, \ref{fig:pmd_all_results})

\begin{table*}[h!]
    \centering
    \setlength\tabcolsep{6pt}
    \caption{Hyperparameters of LFADS and TNDM (adapted from \cite{pandarinath2018inferring}). \\ 'N' - number of units in generator (irrelevant generator for TNDM). 'rel N' - number of units in relevant generator. 'g0' - initial conditions (irrelevant initial conditions for TNDM). 'rel g0' - relevant initial conditions. 'E' - encoder. 'G' - decoder (irrelevant decoder for TNDM). 'rel G' - relevant decoder. '$\lambda_b$' - weight for behaviour likelihood. '$\lambda_Q$' - weight for disentanglement loss. 'KP' - keep probability for dropout. 'B' - batch size.}
    \label{table:hyperparams}
    \vspace*{3px}
    \begin{tabular}{|c|c|c|c|c|c|c|c|c|c|c|c|}
    \hline
    Model & N & rel N& g0 E dim & rel g0 E dim & G L2 & rel G L2 & KP & $\lambda_b$ & $\lambda_Q$ & B \\
    \hline
    LFADS & 64 & N/A & 64 & 64 & 2000 & N/A & .95  & N/A  & N/A & 10   \\
    \hline
    TNDM & 64 & 64 & 64 & 64  & 2000  & 2000 & .85  & .2  & 1000 & 16  \\
    \hline
    \end{tabular}
\end{table*}

\clearpage

\section{Ablation study of disentanglement penalty}
\begin{figure}[ht]
\centering
  \includegraphics[scale=.55]{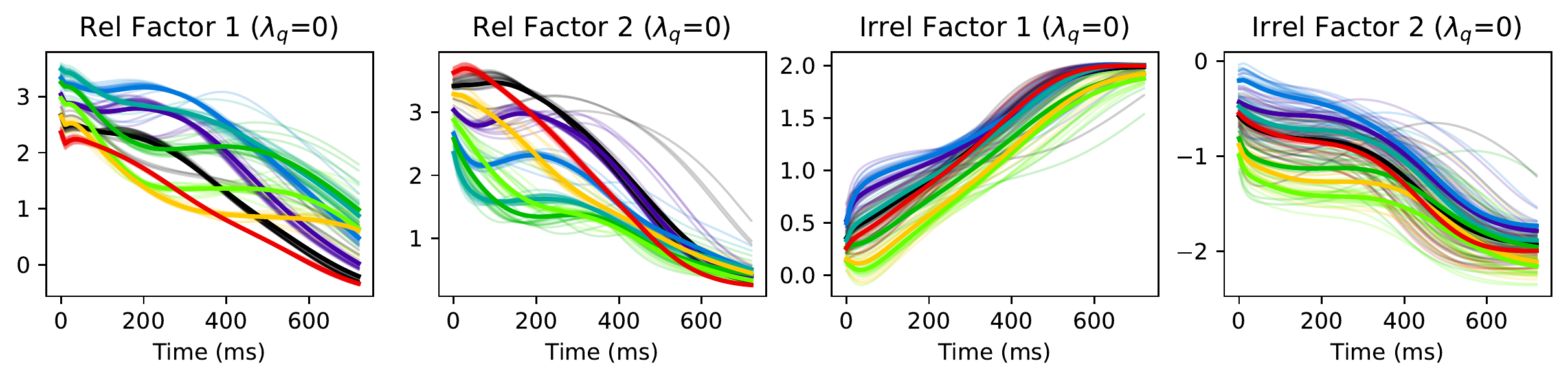}
  \includegraphics[scale=.55]{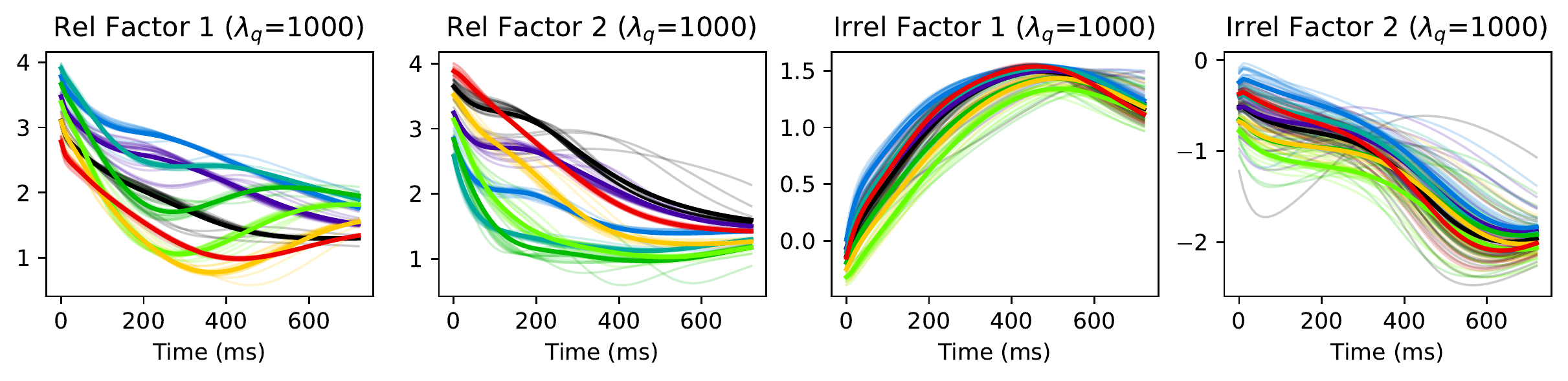}
\caption{Visualizations of the relevant and irrelevant latents for two runs of TNDM with and without the disentanglement penalty. In each box, we plot the condition-averaged inferred latent dynamical factors (along with the single-trial factors). On top, TNDM is applied to the neural data and behaviour with no disentanglement penalty. On bottom, TNDM is applied to the neural data and behaviour with a disentanglement penalty that is weighted by $\lambda_Q=1000$. As can be seen, when the disentanglement penalty is applied, the irrelevant factors contain less behavioural information (condition-averaged irrelevant latents are less separated). Also, the relevant factors seem to share less information with the irrelevant factors.}
\label{fig:disentanglement_ablation_4_2}
\end{figure}

The primary aim of TNDM is to disentangle the behaviourally relevant and the behaviourally irrelevant latent dynamics underlying neural activity. Although there will naturally be some separation between the two sets of factors in TNDM (since the behaviourally relevant factors must reconstruct the behaviour), the factors may still share information. To further encourage disentanglement of the relevant and irrelevant factors, we introduce a disentanglement penalty on the initial condition distributions (described in Section 3 of the main text). Although we did not confirm the efficacy of this penalty exhaustively, in Figure \ref{fig:disentanglement_ablation_4_2} we show TNDM with and without the disentanglement penalty for a specific example. As can be seen, with 2 relevant and 2 irrelevant factors, the penalty forces the irrelevant factors to encode less information about the behaviour and the relevant factors to encode less behaviourally irrelevant information. This is quantified in Table \ref{table:disentanglement_table}. Although there is some evidence that this disentanglement penalty is useful, there is room for improvement as it is only applied to the initial condition distributions and not the factors themselves.

\begin{table*}[ht]
    \centering
    \setlength\tabcolsep{6pt}
    \caption{Behavioural prediction using the relevant and irrelevant factors shown in Figure \ref{fig:disentanglement_ablation_4_2}. To determine how much behavioural information is stored in the relevant and irrelevant factors, we regress the inferred training relevant and irrelevant latents (with a linear ridge regression) to the training behaviours . We then report the test $R^2$ of the regression using the test relevant and irrelevant latents and the test behaviours. As can be seen, the irrelevant factors learned by TNDM with no disentanglement penalty contain more behavioural information (higher test $R^2$).}
    \label{table:disentanglement_table}
    \vspace*{3px}
    \begin{tabular}{|c|c|c|c|}
    \hline
    Rel Facs ($\lambda_Q=1000$) & Irrel Facs ($\lambda_Q=1000$) & Rel Facs ($\lambda_Q=0$) & Irrel Facs ($\lambda_Q=0$) \\
    \hline
    0.890 & 0.245 & .883 & 0.373 \\
    \hline
    \end{tabular}
\end{table*}

\clearpage
\section{Preferential subspace identification failure mode}

In this supplement, we perform two experiments that provide evidence that PSID does not learn latent dynamics that are informative about the neural activity for our M1 dataset. As there are very few adjustable hyperparameters for PSID, we only had to set the number of block-rows in  (i.e. "future horizon" and "past horizon") and the smoothing method for the spiking data. For the horizon, we set the value to the default of 10 (although changing this parameter seemed to have little effect) and for the smoothing we used a Gaussian filter with a standard deviation of 50 ms. We perform our analysis of PSID with 2 behaviourally relevant factors. All code for these analyses can be found at \href{https://github.com/HennigLab/psid_technical_report}{https://github.com/HennigLab/psid\_technical\_report}.

\begin{figure}[ht]
\centering
  \includegraphics[scale=.3]{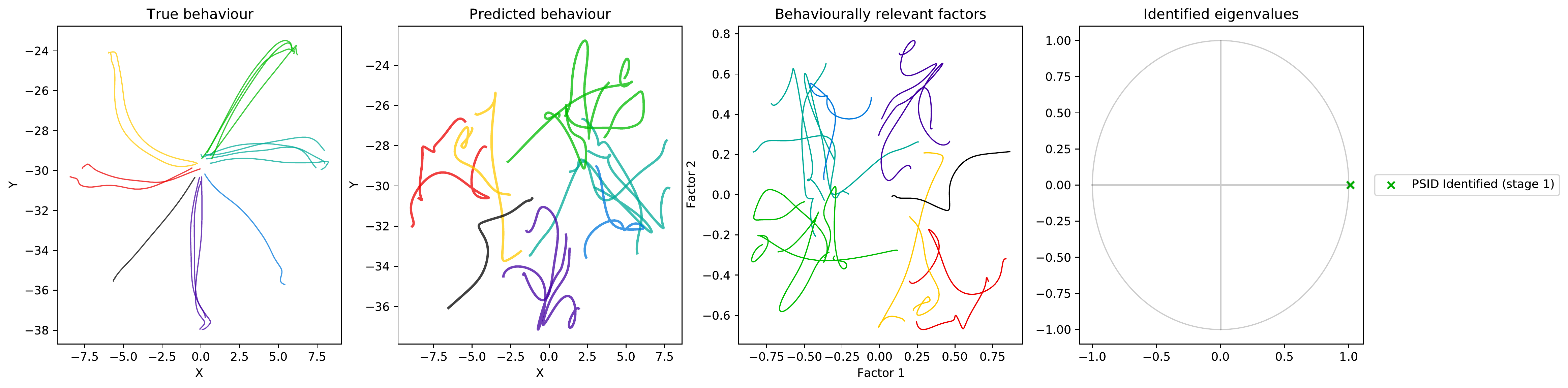}
  \includegraphics[scale=.3]{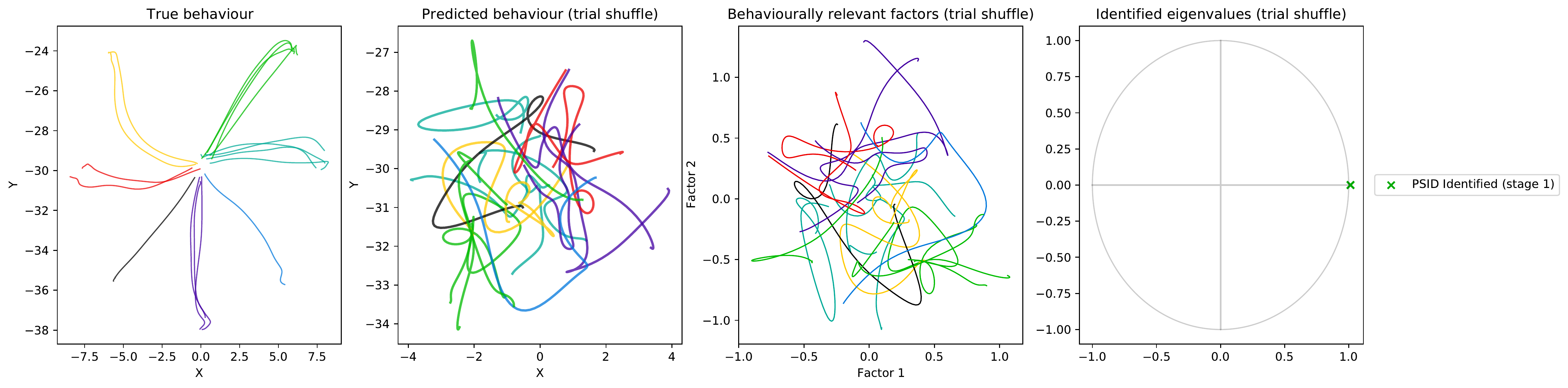}
  \includegraphics[scale=.3]{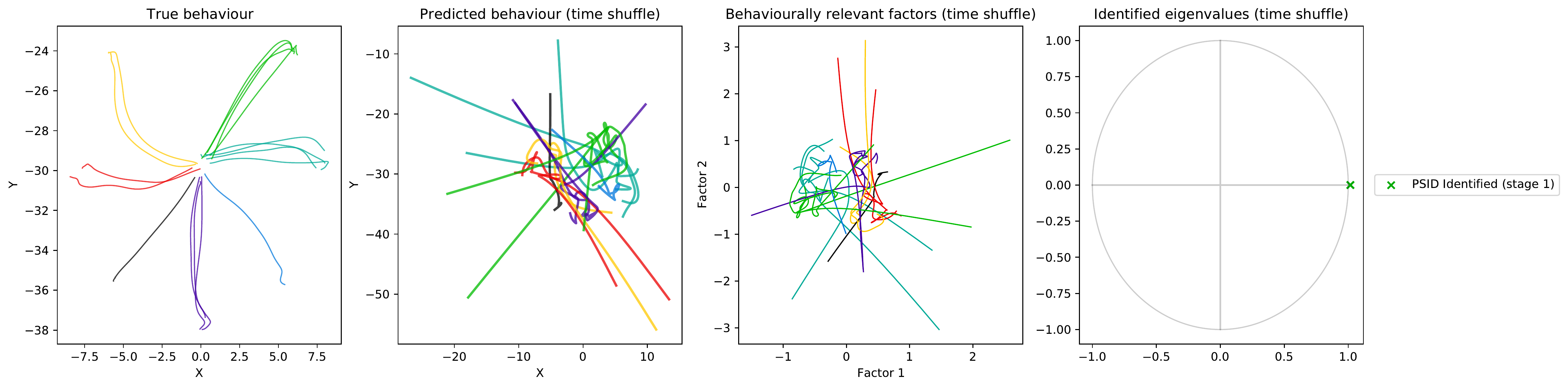}
\caption{Visualizations of the relevant latents and behavioural predictions of PSID (using a Kalman filter prediction approach) on regular and shuffled neural data. On top, visualizations are shown from PSID when trained on the neural data and behaviour normally. The behaviour prediction resembles random walks and the $A$ matrix is close to the identity. In the middle and the bottom plots, visualizations are shown from PSID when trained on shuffled neural data (by trial and in time) and behaviour. While the behaviour prediction is significantly worse, the $A$ matrix is again close to the identity. These experiments suggest that PSID is finding latent dynamics that are uninformative about the neural activity.}
\label{fig:psid}
\end{figure}

The first experiment we run is training PSID normally on the center-out reach dataset and then inspecting the learned state-transition matrix A. According to the \citet{saniModelingBehaviorallyRelevant2020}, the state-transition matrix $A$ "[characterizes] the behaviorally relevant neural dynamics". As can be seen in Table \ref{table:psid_eigen} and Figure \ref{fig:psid}, the learned state-transition matrix $A$ is approximately the identity with eigenvalues that have real value 1 and an insignificant complex component. To better understand if the state-transition matrix $A$ being an identity matrix still meaningfully characterizes the neural activity, we also train PSID on time shuffled and trial shuffled neural data. In both cases, the state-transition matrix $A$ is again approximately the identity matrix with an insignificant complex component. These experiments suggest that the learned identity matrix is not informative about the neural activity for PSID.

\begin{figure}[ht]
\centering
  \includegraphics[scale=.6]{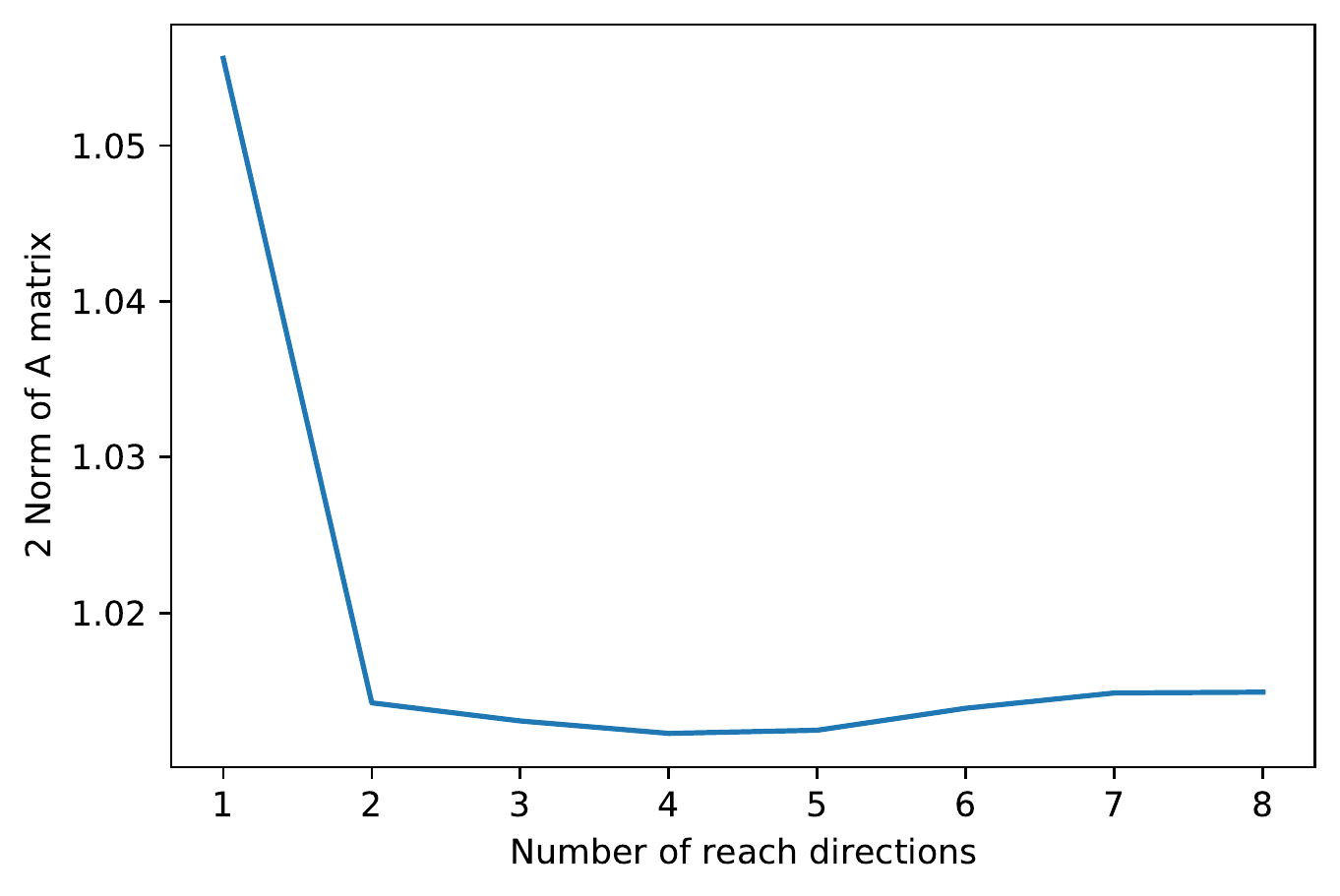}
\caption{Visualization of the 2 norm of the state-transition matrix $A$ when PSID is trained on 1 to 8 different reach directions. As the number of reach directions increases past 1, the 2 norm quickly collapses to 1 which implies that the A matrix is close to the identity. This experiment suggests that the nonlinear multi-reaching behaviour is challenging for PSID to model as it is a linear dynamical model.}
\label{fig:multireachpsid}
\end{figure}

We postulate that the state-transition matrix $A$ is uninformative about the neural activity due to the nonlinear behaviour. The behaviour is nonlinear across all trials due to the 8 different reach directions. To test if this is the case, we train PSID multiple times with 1 to 8 reach directions. As can be seen in Table \ref{table:nonlinear_reach} and Figure \ref{fig:multireachpsid}, the state-transition matrix $A$ matrix quickly collapses to the identity matrix as the number of reach directions increases past 1. This implies that the multi-reaching behaviour is difficult for PSID to model with linear dynamics.

\begin{table*}[ht]
    \centering
    \setlength\tabcolsep{6pt}
    \caption{In this table, we show the eigenvalues of the state-transition matrix A for PSID trained with 2 behaviourally relevant factor and 0 behaviourally irrelevant factors (the lack of behaviourally irrelevant factors should have no affect on the behaviourally relevant factors). When trained normally, with trial shuffled data, and with time shuffled data, the eigenvalues of A is always close to 1 with an insignificant complex component.}
    \label{table:psid_eigen}
    \vspace*{3px}
    \begin{tabular}{|c|c|c|c|}
    \hline
    Experiment & Normal & Trial Shuffled & Time Shuffled \\
    \hline
    Eigenvalues of A & 1.01 + 0.0016j, 1.01 - 0.0016j & 1.01 + 0.0015j 1.01 - 0.0015j & 1.016, 1.018 \\
    \hline
    \end{tabular}
\end{table*}

\begin{table*}[ht]
    \centering
    \setlength\tabcolsep{6pt}
    \caption{In this table, we demonstrate how the state-transition matrix A of PSID approaches the identity matrix as the number of reach directions increases. We again ran this experiment for PSID trained with 2 behaviourally relevant factors and 0 behaviourally irrelevant factors.}
    \label{table:nonlinear_reach}
    \vspace*{3px}
    \begin{tabular}{|c|c|c|c|c|c|c|c|c|}
    \hline
    \# reach directions & 1 & 2 & 3 & 4 & 5 & 6 & 7 & 8 \\
    \hline
    2 norm of A matrix & 1.06 & 1.01 & 1.01	& 1.01 & 1.01 & 1.01	& 1.01 & 1.01 \\
    \hline
    \end{tabular}
\end{table*}

\clearpage
\section{Simulated data results}

To validate TNDM on simulated data, we run both TNDM and LFADS on synthetic spike trains generated from a Lorenz system, a common benchmark for state space models of neural activity. The spike trains are stochastically generated from a 3-dimensional Lorenz system which is partitioned into behaviorally relevant and irrelevant factors. We transform the relevant factors into the behaviours using a linear transformation and then we transform all the factors into the neural firing rates using a separate linear transformation. The neural firing rates are then transformed into a spike train through an exponential nonlinearity and a Poisson random variable. For our analysis, we set the number of neurons to 30, the number of behaviour dimensions to 4, and the number of behaviourally relevant factors to 2 (out of 3). The initial conditions for the Lorenz system are sampled from a Uniform distribution and the behaviour is corrupted with additive noise sampled from a standard Normal distribution. The code we used for generating the synthetic spike trains can be found at: \href{https://github.com/HennigLab/tndm/blob/main/tndm/lorenz/lorenz_generator.py}{https://github.com/HennigLab/tndm/blob/main/tndm/lorenz/lorenz\_generator.py}.

For training, we evaluate the performance of TNDM and LFADS when trained with 50, 100, and 200 trials. Each trial consists of a single initial condition that is evolved into the three latent factors and then into the behaviour and neural activity. We also evaluate each model across three baseline neural firing rates: 5, 10, and 15 Hz. The results are summarized in Table \ref{table:lorenz}. TNDM is competitive with LFADS across all numbers of trials and baseline firing rates. There is evidence that TNDM outperforms LFADS on the lowest firing rate trials, however, we did not perform an exhaustive hyperparameter search so it is likely that the hyperparameters of LFADS can be adjusted to obtain better results in these cases.

For this analysis, as the relationship between the latent factors and the behaviour has no time lag, we utilize a behaviour decoder for TNDM that has no time lag. This is in contrast to the decoder that we use on real data which allows for capturing arbitrary lags. Also, we utilize a Tensorflow2 re-implementation of the original TNDM and LFADs model for this analysis. These re-implementations can be found at the following repository: \href{https://github.com/HennigLab/tndm}{https://github.com/HennigLab/tndm}. We are currently working on improving and extending this implementation so the commit that should reproduce this analysis is 58a0a71b529f5fbe72ce3f6516daed83ce5885ca.

\begin{table*}[ht]
    \centering
    \setlength\tabcolsep{6pt}
    \caption{In this table, we report the results of TNDM and LFADS when run on the synthetic Lorenz system data. The average $R^2$ of TNDM and LFADS is reported for 3 runs of each model on each of the training conditions.}
    \label{table:lorenz}
    \vspace*{3px}
    \begin{tabular}{|c|c|c|c|c|c|c|c|c|c|}
    \hline
    Firing Rate & 5 &  &  & 10 &  &  & 15 &  &   \\
    Train Trials & 50 & 100 & 200 & 50 & 100 & 200 & 50 & 100 & 200 \\
    \hline
    LFADS & .52 & .86 & .88 & .54 & .88 & .92 & .50 &.87 &.92 \\
    \hline
    TNDM & .71 & .83 & .88 & .66 & .86 & .92 & .67 & .86 & .93 \\
    \hline
    \end{tabular}
\end{table*}

\clearpage
\section{Leave one direction out results}

\begin{figure}[h]
\centering 
  \includegraphics[scale=1]{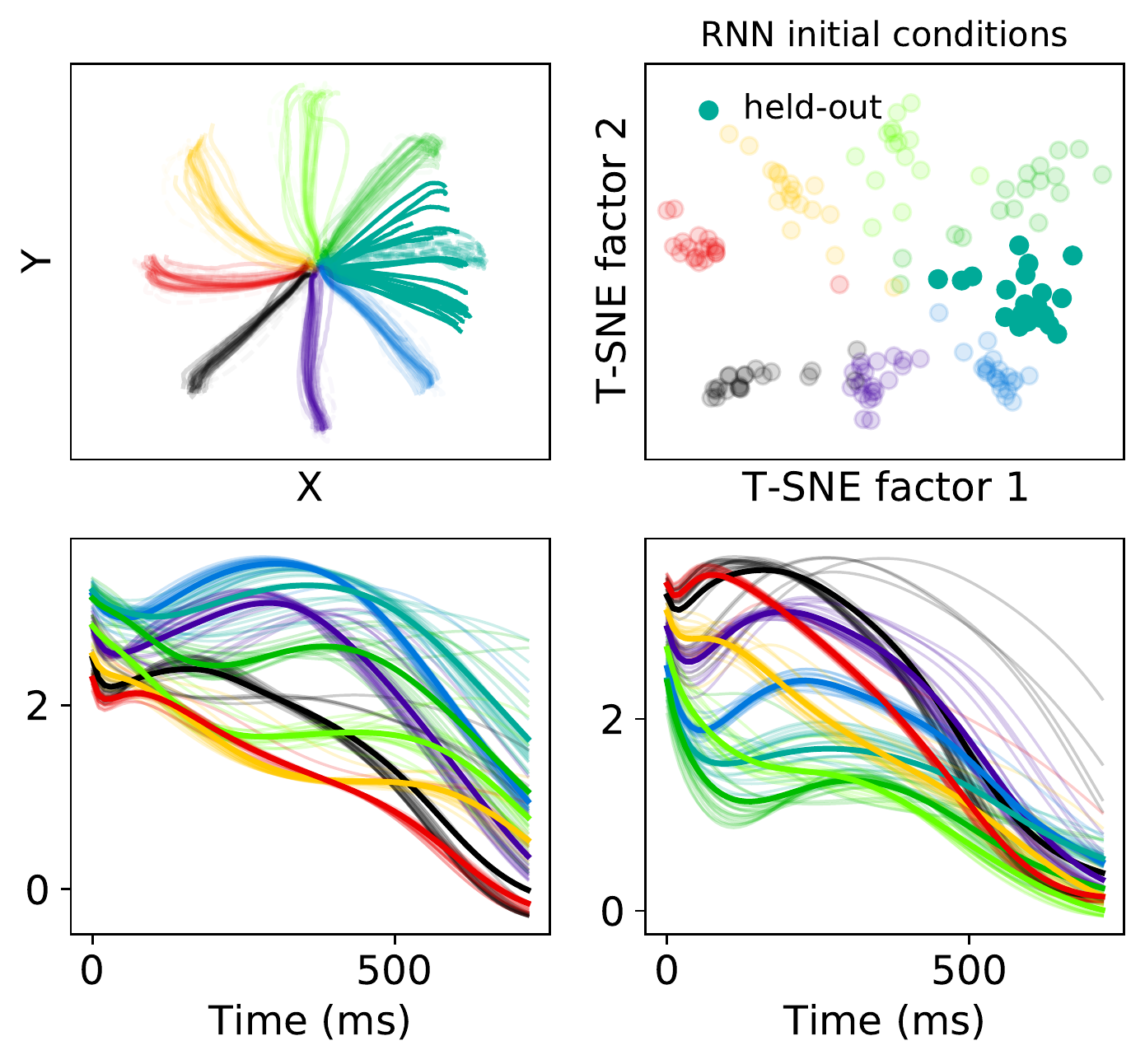}
\caption{We train TNDM on 7 out of the 8 reach directions and then we see if we can infer the initial conditions, latent factors, and behaviour of the held-out reach condition. TNDM is able to recover these fairly well despite the small number of training trials ($\sim$100 trials).}
\label{fig:held-out}
\end{figure}

To illustrate that TNDM learns an interpretable and meaningful latent space, we run an experiment where we leave out one reach direction during training. We train TNDM on 7 out of the 8 reach directions and then we see if we can infer the initial conditions, latent factors, and behaviour of the held-out reach condition. The results are shown in Figure \ref{fig:held-out}. Although not perfect, TNDM recovers initial conditions, latent factors, and behaviours for the held-out reach condition that are close to the model that is trained with all 8 reach conditions. This suggests that TNDM is able to learn latent dynamics that meaningfully capture the behavioural/neural manifold of reach.  We imagine this result can be improved with more trials ($\sim$100 trials is quite limited) and with more hyperparameter tuning. 

\end{document}